# Graphical Prediction of Trapped Mode Resonances in Sub-mm and THz Waveguide Networks

Matthew A. Morgan – matt.morgan@nrao.edu
Shing-Kuo Pan – span@nrao.edu

National Radio Astronomy Observatory
1180 Boxwood Estate Rd.
Charlottesville, VA. 22903

**Abstract**

An analytical method for the visualization and prediction of trapped-mode resonances based on the dimensions of a dispersive microwave network is described. The method as explained is intuitive, easy to implement, and has proven itself to be a useful tool in the avoidance of problems associated with trapped modes prior to fabrication, as well as to correct those problems in designs where this design detail was overlooked.

## I. Introduction

Whenever the waveguides in a network are overmoded, there exists the possibility of trapped modes – that is, modes which are free to propagate within some part of the structure but have no means of propagating outside of it, nor dissipating their energy internally (except, typically, by the nearly-vanishing Ohmic loss of the metal walls). This is especially common with Orthomode Transducers (OMTs) where the required operating bandwidth makes multiple spurious modes difficult (but not impossible [1]) to avoid. The most common symptom of trapped modes is the appearance of a suck-out – a sharp, isolated notch in insertion loss – somewhere in the pass-band of the system. For many applications, including some in radio astronomy wherein wideband OMTs are standard components and sharp spectral features may be mistaken for real signals, the presence of a suck-out is unacceptable.

Yet, despite a general familiarity with the causes and effects of trapped mode resonances, there is no common practice for their analytical prediction prior to cutting metal. Simulated models, which are often identically lossless and possessed of mathematically-perfect symmetry, do not always reveal them. As such, mode-resonances are most often discovered for the first time during initial testing of a prototype. Then, without a reliable analytical method to guide a revised design, one usually resorts to inserting pieces of absorber to dampen the trapped mode cavity, or placing tighter constraints on manufacturing symmetry to avoid excitation of the unwanted modes. The former may degrade performance, while the latter will almost certainly reduce yields, especially at higher frequencies where tolerances cannot be held as tight. At sub-millimeter wavelengths, the required symmetry is practically impossible to achieve. In addition, field distortion due to losses in the walls near the cutoff of a spurious mode, which is most severe in the sub-mm-wave regime, can also induce mode conversion and provide the linkage between the dominant and trapped modes.

The authors have adopted a graphical approach, which not only helps to visualize potential problems prior to fabrication, but also to suggest easy solutions for relatively mature designs that exhibit resonances. The underlying principles and implementation of this method are described in this article.

## II. The Cause of Trapped-Mode Resonances

In order for a suck-out to develop, three conditions must be satisfied:
1. A spurious mode must propagate.
2. The mode must be trapped.
3. The size of the trap must be sufficient for the mode to become resonant.

The first item in the above list also suggests the first line of defense against trapped mode resonances from occurring – simply ensure that the waveguides do not support unwanted modes in the passband. Despite conventional wisdom, this is possible even for systems approaching very wide relative bandwidths [1]-[2]. Such spurious-mode-free designs may ultimately offer the best reliability and production yield of any wideband OMT topology at sub-millimeter-wave frequencies.

When spurious modes are unavoidable, one should take steps to ensure that the second condition above is not met. It can be surprising how easy it is for a high-order mode to become trapped. Simple visual inspection is usually not sufficient, as two waveguides that appear to be similar in size, but different in shape, may still have significantly different cutoff frequencies, and the cutoffs of the higher-order modes do not always follow the same trends as the dominant modes. Consider, for example, the sub-millimeter wave circular-to-square transition shown in Fig. 1. Although from this side-view the waveguide sections appear to taper monotonically down to the right, giving one the impression that any mode within is free to escape to the left, such is not the case. In fact, we will show later that there is a mode trapped in the first three square waveguide sections that would lead to a suck-out at approximately 498 GHz.

When all else fails, faulty designs that exhibit a trapped-mode resonance may be repaired by close consideration of the third condition above. Once again contrary to the common assumption, it is not the case that all trapped modes lead to resonances. Modes can only be trapped over a finite range of frequencies, and they need not become resonant at any point within that range. It is therefore not necessary to tune a resonance completely out of the operating band – even suck-outs occurring near the center-frequency of the system can be made to vanish entirely by relatively minor changes in a few key dimensions. Specific examples of this will be given later.

Notably absent from the list at the beginning of this section is that most spurious modes also require some kind of discontinuity or asymmetry to be coupled with the dominant modes. While strictly true in a theoretical sense, the authors contend that truly perfect symmetry is an impossible ideal that can only be achieved in simulation. In practice, and especially at sub-millimeter-wave frequencies, there is always some asymmetry present, if only as a consequence of imperfect manufacturing or part-to-part alignment. Further, the development of a suck-out is intrinsically non-graceful – the onset of bad effects is practically instantaneous on either side of perfect symmetry. It is far better to assume that if a trapped mode can propagate, it will propagate. Fine-tuning the symmetry only reduces the coupling into the trapped mode, increasing the loaded $Q$ of the resonator. While such efforts may reduce the suck-out below the noise floor in a laboratory environment, it will always be present to some degree.

### III. Graphical Representation of Trapped Mode Resonances

The three conditions for mode resonance introduced in the previous section can be visualized graphically as shown in Fig. 2, wherein the cutoff frequency of a spurious mode is plotted as a function of position along the signal path in a waveguide network. Since waves will propagate in this mode at any frequency above this curve, the first condition is met simply if the line at any point passes below the upper frequency limit of the operating band for the system in question. The second condition is then met if at any point the curve reaches a local minimum – a *well*, where the mode can propagate over certain frequencies that are cutoff at both ends. Finally, the third condition is met if the well is deep enough and/or long enough that the cavity is half of a guided-wavelength long at a frequency where the mode is still trapped. Additional harmonic resonances then occur at integer multiples of a half wavelength up to the point where the mode is no longer cutoff at one end or the other.

To be rigorous, the exact resonant frequencies will depend on the reactive impedances presented by the discontinuities at both ends, as well as intermediate reflections from internal geometric and impedance changes, but the half-wave condition is a good approximation in most cases. This may be estimated by integrating the propagation constant across the cavity to find the total phase

TABLE I: DOMINANT MODE COUPLING PAIRS
FOR CIRCULAR AND RECTANGULAR WAVEGUIDES

| Circular Waveguide Labels | Square/Rectangular Waveguide Labels | Circular Waveguide Labels | Square/Rectangular Waveguide Labels |
|---|---|---|---|
| $TE_{11a}$ | $TE_{10}$ | $TE_{31b}$ | $TE_{21}$ |
| $TE_{11b}$ | $TE_{01}$ | $TM_{21a}$ | $TM_{22}$ |
| $TM_{01}$ | $TM_{11}$ | $TM_{21b}$ | $TM_{31}$-$TM_{13}$ |
| $TE_{21a}$ | $TE_{11}$ | $TE_{41a}$ | $TE_{31}$-$TE_{13}$ |
| $TE_{21b}$ | $TE_{20}$+ $TE_{02}$ | $TE_{41b}$ | $TE_{22}$ |
| $TM_{11a}$ | $TM_{21}$ | $TE_{12a}$ | $TE_{32}$ |
| $TM_{11b}$ | $TM_{12}$ | $TE_{12b}$ | $TE_{23}$ |
| $TE_{01}$ | $TE_{20}$-$TE_{02}$ | $TM_{02}$ | $TM_{31}$+$TM_{13}$ |
| $TE_{31a}$ | $TE_{12}$ | $TE_{22a}$ | $TE_{31}$+$TE_{13}$ |

$$\theta(f) \approx \int_{x_1}^{x_2} \beta(x)dx = \int_{x_1}^{x_2} \sqrt{k^2 - k_c^2(x)}dx$$
$$= 2\pi\sqrt{\mu\varepsilon}\int_{x_1}^{x_2} \sqrt{f^2 - f_c^2(x)}dx \quad (1)$$

and represented graphically by shading in the bottom of the well.

It is noteworthy that this derivation parallels similar approximate methods used to solve the Riccati equation for waveguide tapers (and more generally in solving problems in other fields such as mechanics, optics, acoustics, etc.) wherein longitudinal changes are assumed to be gradual so that the local transmission properties are approximated by those of a uniform line having the same cross-sectional dimensions at a given plane [3]. Eq. (1) may be recognized in this context as an appropriate generalization of the phase accumulation across such a taper in dispersive waveguide.

Note that the form of Eq. (1) suggests the fast-wave dispersion characteristic commonly associated with uniform cylindrical waveguides, but it applies to non-dispersive transmission lines as well, such as coaxial cable or stripline, with $f_c$=0 substituted for the cutoff frequency. It would have to be modified to capture the dispersion characteristics of some more complicated waveguides, however, such as slow-wave or other periodic structures.

The resonance condition is satisfied when $\theta=N\pi$ for any positive integer $N$. The method thus consists of plotting the cutoff frequencies of all the lowest order modes that can propagate in the structure through each branch of the waveguide network, examining these plots to identify local minima, and then evaluating the integral (1) for each of the minima to determine if the trapped mode will resonate. No assumption is made about how the trapped modes are excited. As stated above, it is simply assumed that if a mode can propagate, then it will be present at some level.

Before going into specific examples of this technique in action, a few words of caution are warranted. First, the idea that a single mode can pass, uninterrupted, from one waveguide to another having a different size and shape is a bit of an approximation. In a rigorous sense, when two differently-shaped waveguides are brought together, each mode from one side couples into an infinite number of (possibly evanescent) modes on the other. A full mode-matching solution [4] would be required at each interface to be theoretically exact. At normal operating frequencies, however, the vast majority of the coupled modes in the infinite series will be cutoff. In practice, since a mode can only be trapped if *none* of its counterparts can propagate, the dominant coupling is the only one that matters here, and a simple one-to-one pairing makes sense. In most cases, the dominant mode pairs across two waveguides will be obvious from appearance (but not by their conventional labels), as in the first row of Fig. 3. When degenerate modes are involved, however, it may be necessary to consider super-positions, as illustrated in the second and third rows of Fig. 3. Table I summarizes the dominant mode pairings for the most common cases involving circular, square, and rectangular waveguides.

Note that it is rarely necessary to extend the analysis beyond the first five or six modes in any waveguide. Not only do these modes cutoff at high frequencies that are usually well beyond the operating band, they are also increasingly unlikely to become trapped. A $10^{th}$-order mode, for example, would be difficult to trap except by pathological design, as it would couple into so many other modes of lower-order than itself that almost certainly one of them can propagate away. In the high-order limit, the increasingly dense cutoff frequencies blend together into a continuum of possible modes that permit easy coupling from one to another with vanishingly small perturbation to the overall electromagnetic field.

In any case, it should be clear from Fig. 3 that the standard subscript notation for $TE_{mn}$ and $TM_{mn}$ waves is a poor way to keep track of modes in this analysis, as the indexing is directly connected to the coordinate system in which the field solution was derived and may differ from one waveguide section to the next (assuming such closed-form solutions are even possible). For simplicity, the authors will apply the circular-waveguide labels to the entire chain of modes into which they couple, regardless of the actual geometry of the waveguide at a given position.

## IV. Analysis Examples and Measurements

### A. Circular to Square/Rectangular Step Transitions

To illustrate this method, we return to the circular-to-square waveguide transition shown in Fig. 1. The cutoff frequencies for these waveguides are readily available from theory, and are plotted as a function of position in Fig. 4. Although the simulated return and insertion loss of this structure is very good, the analysis described herein reveals a problem. Note that the (circular) $TE_{21,a}$ mode cutoff drops precipitously at the circular-to-square junction 0.13 inches into the device, far more so than the dominant mode cutoff does. This leads to a trapped-mode well that is deep enough to become resonant at 498.8 GHz, as determined by (1). As a consequence of this result, the design of the transition was changed to correct the error prior to fabrication.

The circular-to-square transition just described is the linkage between the feedhorn and OMT in a full dual-polarization receiver. A parallel development on the same program is a direct circular-to-rectangular transition (bypassing the OMT) designed for test and trouble-shooting purposes. A detail from the manufacturing drawing of this component is shown in Fig. 5. The cutoff frequency spectrum was calculated and plotted in the same manner as before, with the result shown in Fig. 6. In this case, there is a trapped mode, $TE_{21,a}$ (recalling that circular waveguide mode labels are used), however that mode never achieves resonance. At the highest frequency at which the mode is trapped, 470 GHz, the electrical length of cavity is only 51 degrees, much less than the 180 degrees that would be required for the first resonance to occur. Below this frequency, the wavelength is longer and the electrical length even smaller. Above this frequency, where the wavelength is shorter, the mode is free to escape. Since the coupling into this mode is very weak by symmetry, and since there is no resonance to magnify it, the effect it would have on the passband should be smooth and negligible. It is only through a resonance that trapped mode effects become sharp and significant. Of course, gross misalignment could cause performance issues associated with higher-order modes as well, but that is true whether or not any such modes are trapped.

### B. ALMA Band 6 Feedhorn-OMT

Another example which was only analyzed this way after fabrication is shown in Fig. 7. This diagram consists of a section of a conical, corrugated feedhorn and Bøifot-style OMT made for Band 6 (211-275 GHz) of the Atacama Large Millimeter Array (ALMA) [5]. The transition from circular-to-square waveguide occurs inside the feedhorn and the flange interface between the horn and the OMT is in square waveguide. A thin, metal septum and vertical wires complete the Bøifot junction. Strictly speaking, the forward tip of the septum creates a very short section of quasi-coaxial waveguide, but this detail is omitted from the analysis for simplicity. The feed-OMT combination is modeled instead as a simple series of circular-, square-, and rectangular-waveguide sections following the side-arm branch of the OMT. The cutoff frequencies are plotted in Fig. 8.

Two trapped-mode resonances are predicted from this plot. The first, at 227.5 GHz, arises from the $TE_{21,a}$ mode trapped between the circular-to-square junction in the feedhorn and the Bøifot junction in the OMT. This resonance is particularly insidious because it requires the two components to be mated to manifest itself, and would be sensitive to (among other things) the alignment of the flanges at the interface. Measurements of either component alone may not even show a resonance, and if they did, it could be at an entirely different frequency. This fact highlights the need to consider whole systems, not just sub-components, and to be especially careful of interfaces that occur in overmoded waveguide – or better yet, to avoid such interfaces altogether.

The second resonance, predicted at 256.5 GHz, occurs in the passive combiner between the two side-arms. The two rectangular waveguides are first joined to form a square waveguide, then the square waveguide is tapered back down to rectangular again. This causes the modes in the combiner to drop in cutoff frequency, then rise again, forming a cutoff well. As just seen, such a feature is a good recipe for trapped-mode resonances to occur. Both the $TE_{21,a}$ and $TM_{01}$ modes, which are degenerate in square waveguide, exhibit this resonance.

Obviously, there are many other cutoff wells in this plot, however the rest are either too high in frequency (thus out-of-band) or too short to support a resonance. The corrugations in the feedhorn, for example, create many dips and peaks in the plot on the left, but each is much less than half a wavelength long at the frequencies of interest.

To be fair, the potential for trapped modes was well understood by the designer of this OMT [5] and the tradeoffs associated with the sidearm combiner were well documented. Nevertheless, the issues were not fully appreciated by the project at large until most of the production feedhorns and OMTs had been fabricated.

Although the occurrence of trapped modes was harmful to the Band 6 project, the situation did present a unique opportunity to verify the theory through measurement. The lower (and, it turns out, more prominent) resonance at 227.5 GHz was not observed during qualification of the prototype components because, as described earlier, it is only present at that frequency when the horn and OMT are joined together. Resonances which were measured at the component-level were dismissed as artifacts of the tapered square-to-rectangular transitions used in the test set – an assumption seemingly confirmed when those resonances disappeared after inserting resistive vanes into the tapers. Of course, since those resistive vanes absorb energy from the higher-order modes as well as the orthogonal dominant mode for which the vanes were intended, the basis for dismissing these resonances was not entirely justified.

Y-factor noise measurements proved to be the most effective way of observing these resonances in actual hardware. Fig. 9 and Fig. 10 show measurements of these resonances in actual ALMA Band 6 receivers. Multiple curves on each plot correspond to overlapping RF coverage with different LO tunings – further verifying that the resonant feature occurs in the RF signal path and not the IF. The first resonance is evident in Fig. 9 at 229.9 GHz, slightly higher than predicted. A small shift upward in frequency, about 0.4%, was expected due to shrinkage of the metal (brass) which is cooled in this application to 4 Kelvin. The discrepancy here is much larger than that, however, and is probably due to the geometric approximations that were made around the septum area to simplify analysis. The cutoff plane for the $TE_{21,a}$ mode must be closer to the tip of the septum than was estimated, effectively shortening the trapped-mode cavity. Nonetheless, an error of only 1.1% relative to prediction (0.7% after accounting for thermal contraction) is an encouraging validation of the theory. The higher predicted resonance at 256.5 GHz was also detected (Fig. 10), although in fewer production units and at a weaker level. In this case, the full geometric detail of the cavity was included in the model, and the predicted location of the resonance was much more accurate.

The impact of trapped mode resonances in spectral line observations using low-noise heterodyne systems is evident from these figures. By their nature, the system frequency response around such features – whether it be in amplitude, phase, beam pattern, polarization properties, or noise – is locally very sensitive to external influences such as vibration and temperature, making it difficult establish a proper receiver baseline and correct for it. The impact on continuum observations may be less significant, as other contributions may tend to dominate over broad bandwidths (e.g. optical imperfections contributing to imperfect polarization isolation). Specific requirements will of course depend on the application, but some useful discussion is found in the context of radio astronomy applications in [6] and [7]. More general treatments of mode conversion in waveguides and feeds are found in [8]-[11].

Because the lower suck-out appears near the bottom of the Band 6 operating range, it was initially suggested that the leading square waveguide in the OMT could be lengthened, thus lowering the resonant frequency and tuning it out of band. However, the mode in question is very near cutoff in this region, and thus has an extremely long guided wavelength, diverging to infinity as the resonant frequency is lowered. Large changes in length have only a small effect on resonant frequency, and no matter how long this section was made, the resonant frequency would never get lower than the cutoff of the upper step in the well at 225.5 GHz, which is still in band. Close examination of Fig. 8 suggests a better solution. With the $TE_{21}$ mode cutting off very near to 230 GHz in the circular throat of the feedhorn, the well is just barely deep enough as it is to support the first resonance. This analysis confirmed that a mere 0.8% increase in the diameter of this one section would be sufficient to eliminate the resonance with a comfortable margin. Interestingly, the suck-out in this way is not tuned out of band, but entirely out of existence, vanishing from the spectrum in place, and with little observable effect on the behavior of the rest of the system. This modification was made in a small number of later feedhorns and the resulting clean spectrum for one such unit is shown in Fig. 9. Several more were tested with equally dramatic improvements in performance.

The second resonance at 256 GHz in this case would likewise be easy to fix. By tapering the side-arm waveguides to half-height prior to combining them instead of after, the higher-order modes can be made to raise and then lower instead of the reverse, avoiding the presence of a cutoff well entirely for a small compromise in insertion loss. However, this modification was not incorporated into any production units.

## C. ALMA Band 8 Feedhorn-OMT

Another real-world scenario to which this analysis was applied is shown in Fig. 11 and Fig. 12. This time, it is the ALMA Band 8 (385-500 GHz) feedhorn and OMT, which contains a modified Bøifot-style junction employing a double-ridged structure instead of septum and polarization wires [12]-[14]. The basic features which lead to trapped modes are the same – the transition in the horn from circular-to-square which blocks higher-order modes from escaping and the combiner of the two side arms which transitions to a tall, almost-square waveguide before tapering down to the usual rectangular aspect ratio. In this case, however, the electrical length of the trap between the horn and the OMT is much larger, enough to support six distinct resonances in the $TM_{01}$ mode, and seventeen in the $TE_{21}$ mode, for a total of twenty-four resonances across the design band.

In this case, no obvious suck-outs in the gain or noise temperature were observed during tests of the production hardware. The design of the OMT cleverly avoids the manual assembly of any portion of the critical mode-separating junction, or the interconnection of any two components not completely defined by numerical machining. It is possible that these measures resulted in a manufacturing precision sufficient to suppress trapped-mode resonances below the noise floor.

What was observed, however, was a systematic and at first inexplicable degradation of cross-polarization performance that spanned almost the entire frequency range of the receiver. This seemed not to be related to either the feedhorn or the OMT, for each would perform well in their individual testing. It wasn't until the full front-end was completed that cross-polarization problems became evident. This phenomenon led to a great deal of consternation until the above analysis provided the first plausible explanation for its behavior. The lack of observable suck-outs, however, made this explanation somewhat difficult to accept. The only measurement that bore out a problem was cross-polarization, but this measurement itself is not typically performed with sufficient frequency resolution to reveal the sharp features one would expect. Eventually, a very fine frequency-scale plot of cross-polarization performance was generated, and is shown in Fig. 13. The characteristic sharp features that span the frequency range, though not strictly proof, are consistent with the trapped-mode analysis. This further emphasizes the principle given in the introduction – that even if the manufacturing tolerance is held to such a high standard as to render trapped mode resonances undetectable by most measurements, they are still necessarily present to some degree, so long as the three conditions stated at the beginning of Section II are met.

It is not clear at this point how much of the degradation shown in Fig. 13 manifests as simple polarization isolation or as a varying cross-polar beam pattern – the former being a purely scalar property of the receiver at each frequency, whereas the latter includes the angular dependence of the feed system (i.e. beam symmetry and the response at large angles). Both are potentially degraded by the presence of higher-order trapped modes in the feed or feed-OMT interface. See [15] for further discussion about the importance of high-order mode excitation in high-performance feeds.

## D. ALMA Band 10 Feedhorn

As a final high-frequency example, we consider a deceptively simple case involving the feedhorn for ALMA Band 10 (787-950 GHz). In this case, there is no OMT at all. Instead, a direct linear taper, shown in Fig. 14 transitions smoothly from the circular output of the feed to a single-mode rectangular waveguide. By appearances, this uncomplicated structure decreases monotonically in size from left to right, creating the perception that the cutoff frequencies of the modes would also be monotonic. Thus, it was thought, no cutoff minima should be present, no modes should be trapped, and no suck-outs should occur.

The above intuition has a flaw, however, namely that while the waveguide is changing in size, it is changing in shape as well. Some modes have lower cutoff frequencies in near-rectangular waveguides than they do in near-circular waveguides, even though they may appear to be smaller. In the case of the $TE_{21}$ mode, as shown in Fig. 15, this causes its cutoff frequency to dip slightly as the taper begins, prior to joining the rest of the modes in the expected positive trend. A very shallow well is created, and the length of the taper is sufficient for this well to support a resonance at approximately 850 GHz. This example illustrates just how misleading one's initial instincts can be and how important it is to consider trapped modes analytically prior to cutting metal on a design.

Still, there is a caveat to this analysis that is illustrated well by this example. The peak cutoff of the $TE_{21a}$ mode at 0.2mm in the plot is only slightly higher than 850 GHz where the cavity is resonant. The mode in this region would be only slightly evanescent, almost propagating. If the evanescent region is sufficiently short, it may allow the mode to "tunnel" out of the trapped mode cavity and escape through the feedhorn to the left of the plot. This highlights another way in which the presented analysis is not totally rigorous. It was stated earlier that the half-wave condition for determining resonant frequency is only approximate and neglects the reactive impedance at the end-points. We now see that the terminating impedance of the cavity may also have a real-part, despite the mode being cutoff in the adjoining section, as a result of the energy that can escape via evanescent mode tunneling. In the authors' view, however, this lack of rigor does not take away from the utility of this analysis as a quick and easy tool for highlighting risk areas that are too often overlooked. A degree of sound judgment is simply required to interpret the results. The fact that the measured performance of the Band 10 receivers has yet to show any evidence of trapped-modes gives some credibility to this line of reasoning. Nevertheless, the application of this technique in the future may prove to be a prudent step in producing more robust designs.

Consider, even if the trapped mode in this taper was sufficiently coupled via evanescent tunneling to the external propagating region, what would be the effect of small changes in the dimensions? As drawn, the region over which the 850 GHz signal is cutoff at the left end of the plot is about 220µm long, or 0.6 wavelengths. If the opening at the end were increased in diameter by only a few microns, the resonant cavity would disappear entirely. On the other hand, a few microns in the other direction and the cutoff length would nearly double. Since evanescent tunneling is *exponentially* dependant on the length of the cutoff region, and the length of this region is clearly very sensitive to minute changes in diameter, the design as shown would be hyper-sensitive to manufacturing tolerance. A much safer design would increase the nominal diameter of the aperture at the left a few percent – a small change which could easily be absorbed into the feed design – to eliminate the potential resonance while leaving some margin for manufacturing errors.

*E. Cm-wave Examples*

Although the emphasis in this article is on sub-mm and THz networks, wherein the *Q*-factors are lower and precise symmetry is most difficult to achieve, this technique has been fruitfully applied to lower-frequency networks as well, where components are typically pushed to broader relative bandwidths and the number of spurious modes that can propagate are increased. Two examples are provided here to illustrate the wide range of structures to which this analysis can easily be applied, in anticipation of the future growth in sub-millimeter-wave technology and its adoption of similar wideband techniques.

The first long-wave example, shown in Fig. 16 and Fig. 17 is an ultra-wideband quad-ridge feedhorn [16] covering approximately 0.5-2.5 GHz. The cut-off frequencies in ridged waveguides depend strongly on the aspect ratio of the ridges and gaps, and it is thus very easy for modes to become trapped, even in physically monotonic tapers. This analysis predicts two potential resonances, one at 560 MHz and the other at 2.5 GHz. Considering the very wide bandwidth of the structure and the number of modes that ultimately can propagate inside it, the presence of only two trapped-mode resonances, both occurring near the band edges, is in fact quite an accomplishment. Note that the dominant mode, represented by the lowest curve on the plot, has a well very similar to $TE_{21a}$, but it is not trapped because the coaxial probes couple the energy from this mode out of the structure and into the receiver.

The last example in Fig. 18 and Fig. 19 consists of a triangular-waveguide-based OMT ([1],[2]) connected to a long circular-to-rectangular taper, similar to that used in ALMA Band 10. Although the triangular OMT itself is totally free of spurious modes over the full operating bandwidth (much less any that are trapped), the long taper used for laboratory testing does propagate high order modes and is long enough for one of them to pass through five distinct resonances. This unavoidably leads to suck-outs that were observed during prototype measurements. In the final application, the taper would be replaced by a feedhorn designed to avoid these resonances. Once again, the plot shows wells in the bottom two curves ($TE_{11a}$ and $TE_{11b}$) which at first glance appear to be trapped, but in fact are not because of the absorption of the coaxial probes that couple into the structure.

## V. Rules of Thumb

A recurring theme in many of the above examples is the transition from circular to square/rectangular waveguide. We have seen that many variations on this common basic structure can lead to trapped mode resonances. It is therefore useful to articulate some rules of thumb to guide an initial design. Three such rules motivated by the examples in this paper are shown in Fig. 20. The first concerns networks where the waveguide passes through a square section on the way from circular to rectangular, as from a feedhorn to an OMT. Regardless of the intervening structure, the $TE_{21}$ mode will be blocked by the rectangular waveguide and is trapped in the square section unless the circle diameter is at least 37.5% larger than the side-length of the square. This ensures that the cutoff of the $TE_{21}$ mode in the circular waveguide is below that of the corresponding mode in the square waveguide.

Even if the first rule is obeyed, intervening sections between the circular and square or the square and rectangular waveguides may yet introduce lower cutoffs and a trapped mode. This was the case in ALMA Band 6, shown by the dip at 5 mm on the *x*-axis of Fig. 8. In the case of a smooth taper between the circular and square section, the circle diameter is required to be about 60% larger than the side of the square to avoid trapping $TE_{21}$.

Finally, if the square section is bypassed and a smooth transition is used between the circular and rectangular waveguides, like that in ALMA Band 10, we require only that the circle diameter is 25% larger than the long side of the rectangle.

These rules of thumb are generally sufficient to avoid problems in feed-OMT assemblies no matter how long the transitions are. It is permissible, however, to violate these constraints, thereby creating a trap, so long as one is prepared to ensure that the electrical lengths are too short for the trap to become resonant.

## VI. Conclusions

The causes and effects of trapped-mode resonances or suck-outs in overmoded waveguide systems have been reviewed in detail. Such resonances introduce rapidly varying system responses (in frequency and potentially in time) that are difficult to calibrate out and can make designs highly sensitive to fabrication tolerance. A graphical technique for visualizing the conditions that lead to trapped-mode resonances has been described, and used to analyze a number of real-world examples in the sub-millimeter-wave regime. The technique was then verified by comparing measured evidence of trapped modes in fabricated hardware against the predictions. Despite a number of simplifying assumptions, the suck-outs were observed within 1% of their predicted frequencies, and disappeared from the data when the corrective modifications suggested by the technique were implemented. These examples highlight the power of this kind of analysis to visualize the problems as well as the simplest solutions to trapped-modes in complex waveguide networks. Even in marginal cases where a trapped-mode is sub-resonant, the technique provides valuable insight into potential problem areas that could be targeted for sensitivity analysis.

Although each of the cases studied was unique, serving to illustrate a wide variety of problems and solutions, there are also common threads from which one can derive some general principles. The most common problem areas, related to transitions between circular, square, and rectangular waveguides, have been condensed into a few practical rules-of-thumb which may guide future designs. Among other lessons learned is that interfaces (flanges) between overmoded waveguides cannot be treated as simply as those between single-mode guides, owing to the highly sensitive nature of trapped-mode resonances which are coupled to the interface alignment. Most importantly, feeds and feed-OMT assemblies having overmoded ports must ultimately be tested in combination with the attachments with which they will be used – individual tests of components using external tapers, whether or not they utilize mode-absorbing vanes, can lead to misleading results.

**Acknowledgment**

The authors wish to thank the many groups in the ALMA project and the Steward Observatory for allowing use of their drawings and measured data for the examples in this paper. Special thanks are also due to Edward Wollack of the NASA Goddard Space Flight Center for insightful discussions. The National Radio Astronomy Observatory is a facility of the National Science Foundation operated under cooperative agreement by Associated Universities, Inc.

**Matthew A. Morgan** (M'99) received the B.S.E.E. degree from the University of Virginia, Charlottesville, in 1999, and the M.S. and Ph.D. degrees in electrical engineering from the California Institute of Technology, Pasadena, in 2001 and 2003, respectively.

He is currently a Scientist and Research Engineer with the National Radio Astronomy Observatory (NRAO), Charlottesville, VA, where he is involved in the design of monolithic millimeter-wave integrated circuits (MMICs) and multi-chip modules (MCMs) for radio telescope arrays. Prior to joining the NRAO, he was an Affiliate of the Jet Propulsion Laboratory (JPL) where he developed MMICs and MCMs for atmospheric radiometers and spacecraft telecommunication systems.

**Shing-Kuo Pan** (S'81-M'85) was born in Taiwan, Republic of China, in 1953. He received the B.A. degree from Fu-Jen University, Taiwan, in 1976, and the M.A., M.Phil. and Ph.D. degrees in physics from Columbia University in 1980, 1981 and 1984, respectively.

Since 1984 he has been with the National Radio Astronomy Observatory, Charlottesville, VA, where he has been involved in research on SIS detectors and in the development of SIS heterodyne receivers for radio astronomy. He is responsible for the development of the Band 3 SIS mixer for the Atacama Large Millimeter Array (ALMA) project, currently under construction by an international partnership between countries in North America (United States and Canada), Europe and East Asia. He also supports the SIS receiver on the Harvard-Smithsonian Sky Survey radio telescope at Cambridge, MA. His other interests include computer-aided design, superconducting device fabrication techniques, and noise mechanisms in superconducting devices.

Dr. Pan is a member of the American Physical Society.

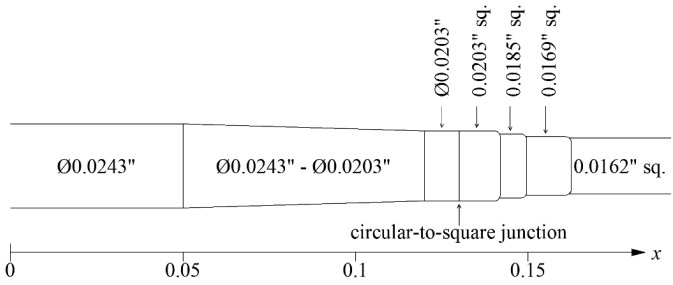

Fig. 1. Circular-to-square waveguide transition. A trapped mode in the highlighted section leads to a suck-out at 498 GHz.

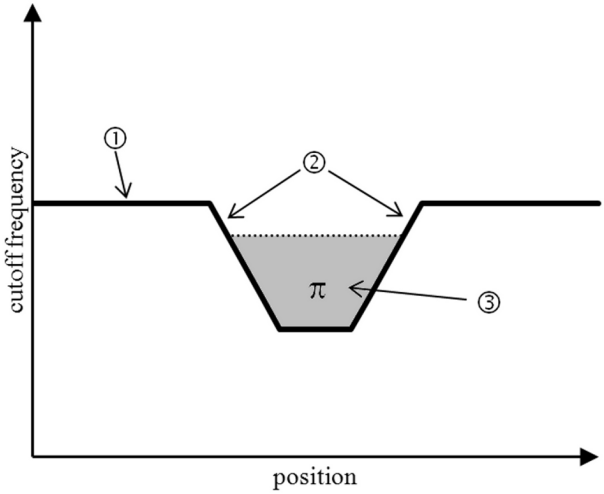

Fig. 2. Graphical representation of a trapped resonant mode.

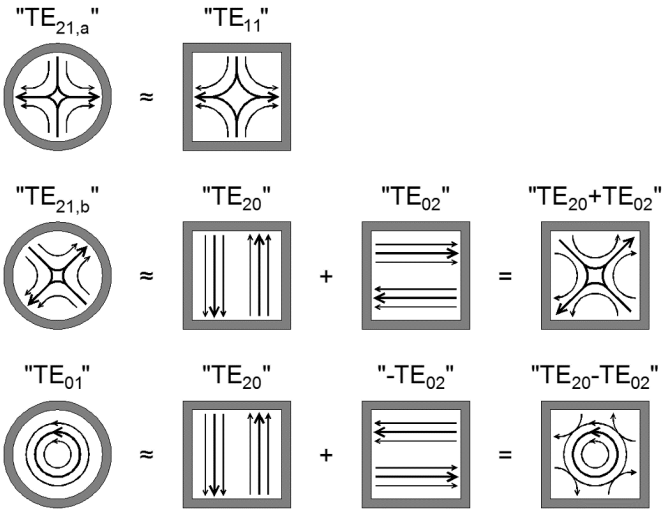

Fig. 3. Illustration of higher-mode equivalence in square and circular waveguides. E-field lines are shown.

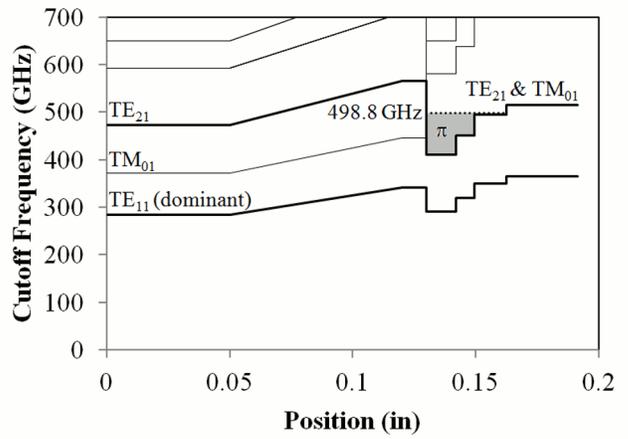

Fig. 4. Plot of the cutoff frequencies of the first few modes in the circular-to-square transition.

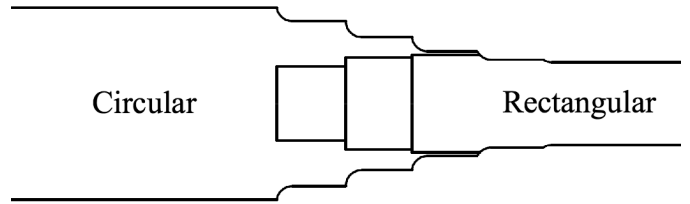

Fig. 5. Detail from manufacturing drawing of a circular-to-rectangular stepped-taper transition.

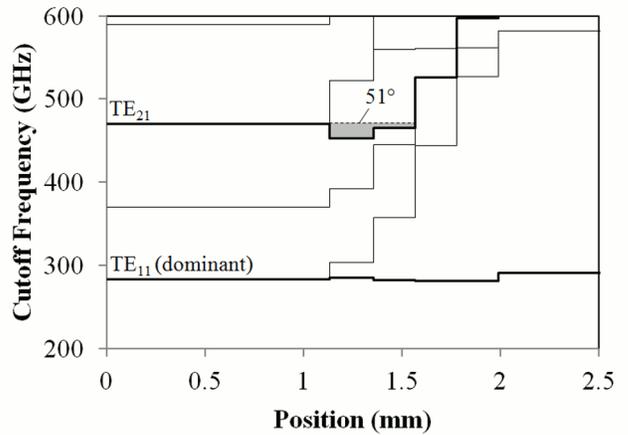

Fig. 6. Mode cutoff plot of circular-to-rectangular stepped-taper transition. The $TE_{21,a}$ mode, though trapped, never becomes resonant.

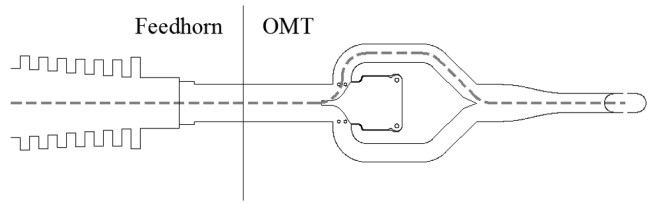

Fig. 7. Section drawing of the ALMA Band 6 (211-275 GHz) feedhorn and OMT. Mode analysis is performed along the dotted line path. Trapped modes are present in the shaded areas.

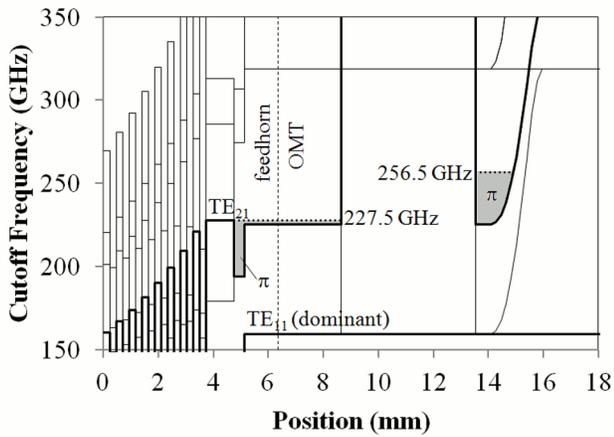

Fig. 8. Plot of the ALMA Band 6 mode analysis predicting two in-band resonances at 227.5 GHz and 256.5 GHz.

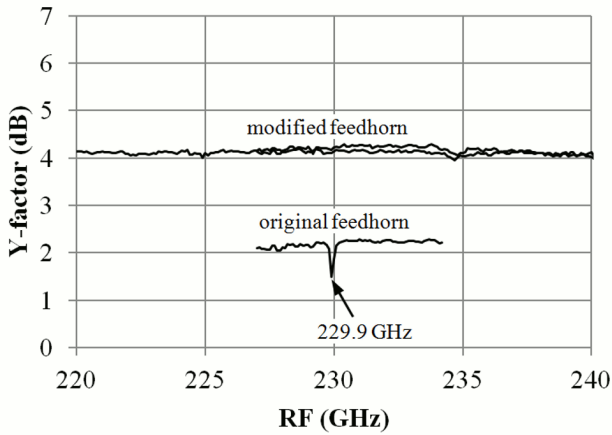

Fig. 9. Measured 230 GHz resonance in an ALMA Band 6 receiver with original and modified feedhorns. Multiple curves correspond to overlapping RF coverage with different LO tunings. Vertical offset added for clarity.

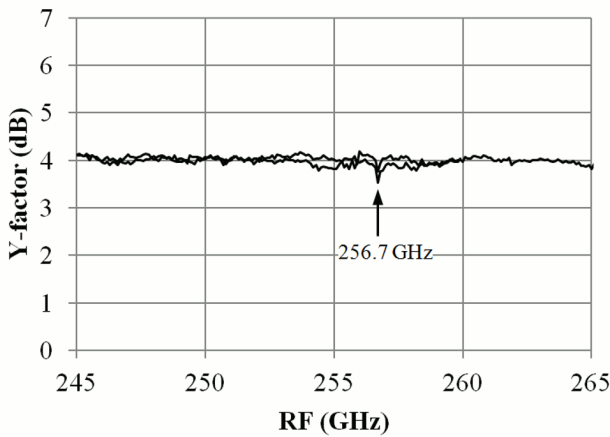

Fig. 10. Measured 256 GHz resonance in an ALMA band 6 receiver. Multiple curves correspond to overlapping RF coverage with different LO tunings.

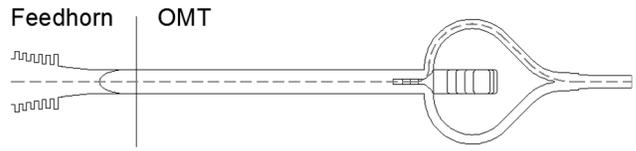

Fig. 11. Section drawing of the ALMA Band 8 (385-500 GHz) feedhorn and OMT. Mode analysis is performed along the dotted line path.

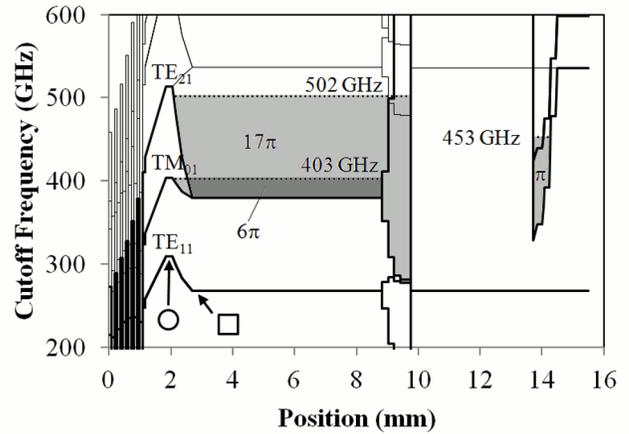

Fig. 12. Plot of the Band 8 mode analysis predicting 24 resonances over the entire operating range of the receiver.

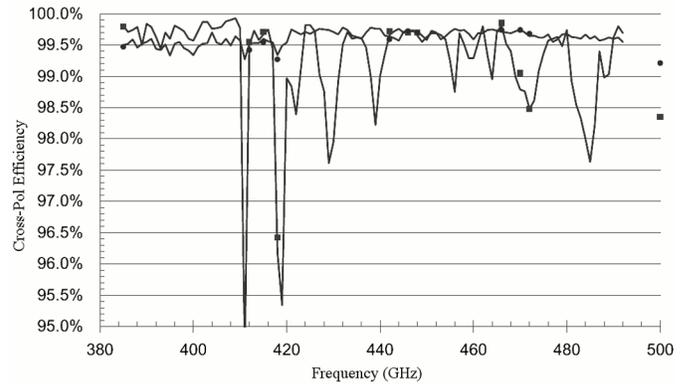

Fig. 13. Fine-resolution cross-polarization measurement of Band 8 receiver showing sharp spectral features. The 2D beam patterns were measured with a sparse grid of 51x51 samples in 1 GHz steps, plotted with solid curves. The marker points were measured more accurately using a fine grid of 101x101 samples.

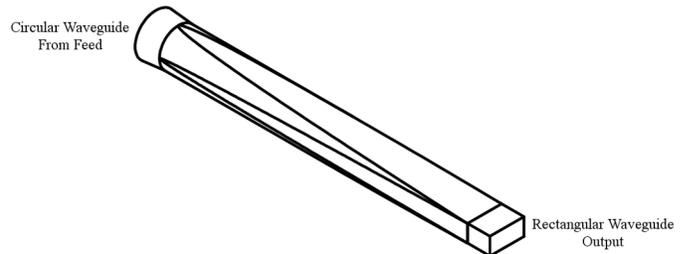

Fig. 14. Smooth taper from the circular waveguide output of the feedhorn to single-mode rectangular waveguide for the ALMA Band 10 (787-950 GHz) receiver.

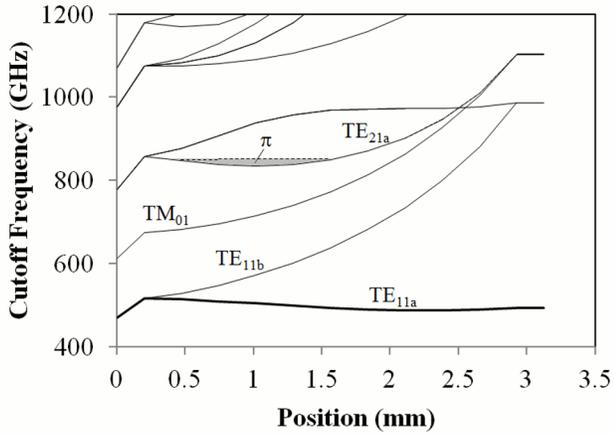

Fig. 15. Plot of the Band 10 mode analysis predicting a trapped mode resonance in the $TE_{21,a}$ mode at approximately 850 GHz.

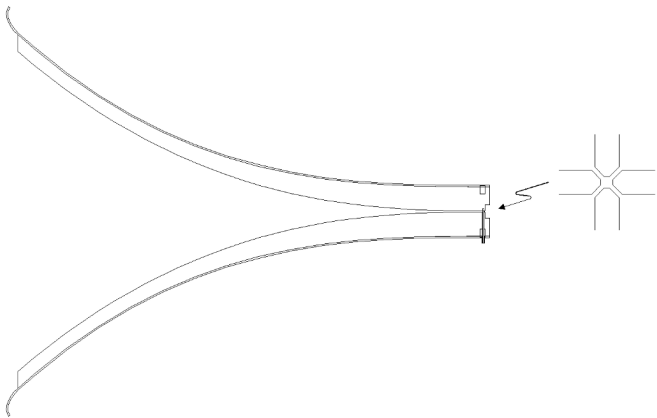

Fig. 16. Diagram of the tapered Quad-Ridge Feedhorn [16].

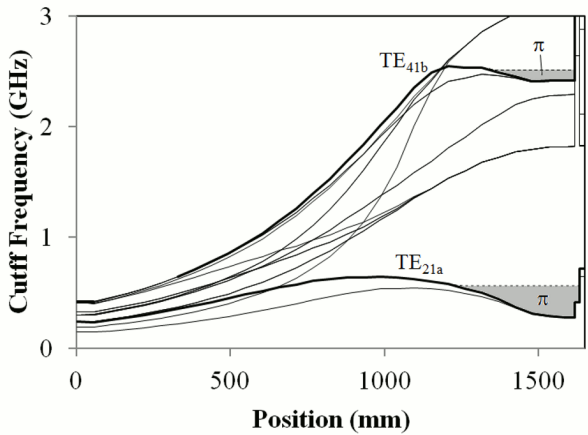

Fig. 17. Mode analysis of the Quad-Ridge Feedhorn predicting resonances at 560 MHz and 2.5 GHz.

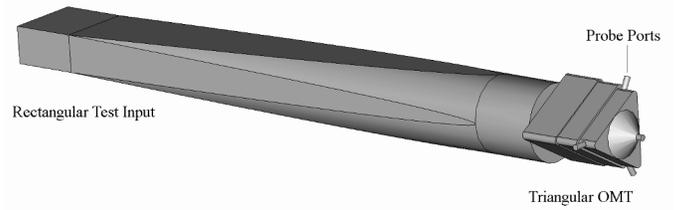

Fig. 18. 3D Model of the S-Band (1.7-2.6 GHz) Triangular OMT test structure.

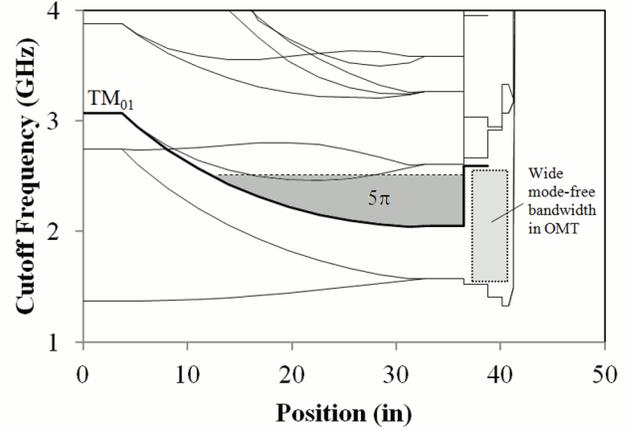

Fig. 19. Mode plot for Triangular OMT test structure.

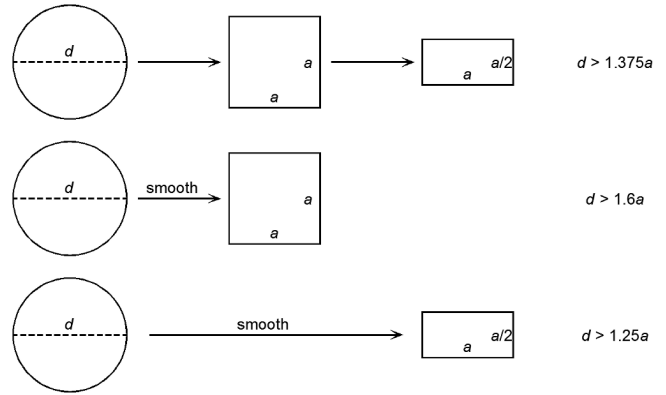

Fig. 20. Rules of thumb regarding circular to square/rectangular waveguide transitions.